\documentclass[conference]{acmsiggraph}

\TOGonlineid{0124}

\usepackage{color}
\usepackage{overpic}
\usepackage{soul}
\usepackage{amsfonts}
\usepackage{wrapfig}
\usepackage{multirow}
\usepackage{subfig}
\usepackage{hyperref}

\newcommand{\mypara}[1]{\vspace{0.05in} \noindent \textbf{#1.}}

\newtheorem{theorem}{Theorem}[section]
\newtheorem{proposition}[theorem]{Proposition}
\newtheorem{definition}[theorem]{Definition}

\newcommand{\vnudge}{\vspace{-.1in}}

\usepackage{amssymb, amsmath, amsfonts}

\graphicspath{{./figures/}}

\usepackage{color}
\definecolor{turquoise}{cmyk}{0.65,0,0.1,0.1}
\definecolor{purple}{rgb}{0.65,0,0.65}
\definecolor{dark_green}{rgb}{0, 0.5, 0}
\definecolor{orange}{rgb}{0.8, 0.2, 0.2}
\definecolor{red}{rgb}{1, 0, 0}

\title{Computational Network Design from Functional Specifications}

\author{Chi-Han Peng\thanks{e-mail:pchihan@asu.edu}\\ Arizona State University / University College London %
\and Niloy J. Mitra\thanks{e-mail:n.mitra@ucl.ac.uk}\\ University College London \linebreak %
\and Fan Bao\\ Arizona State University %
\and Dong-Ming Yan\\ KAUST %
\and Peter Wonka\thanks{e-mail:pwonka@gmail.com}\\ Arizona State University / KAUST }
\pdfauthor{}


\begin{document}

%

 \teaser{
   \includegraphics[width=\textwidth]{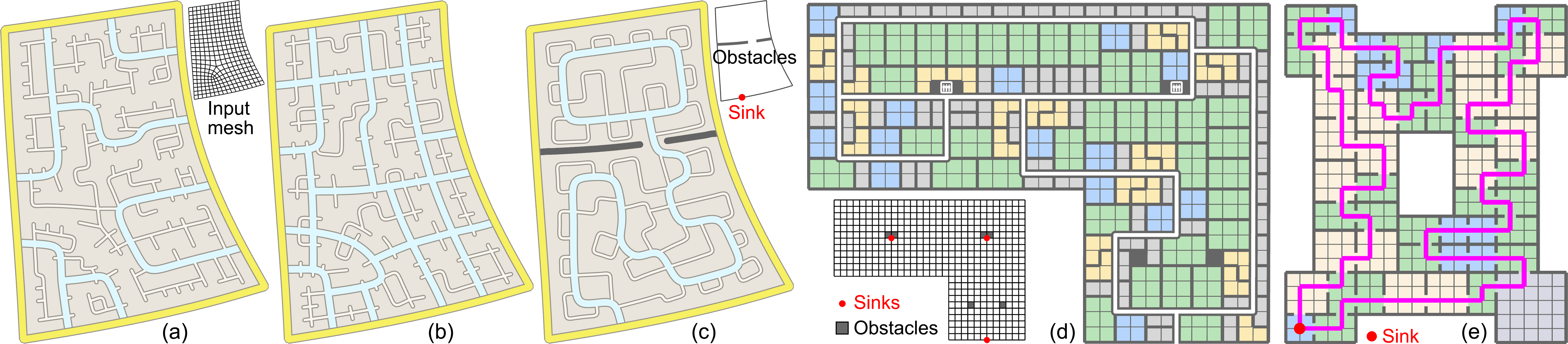}
\caption{We propose an algorithm for generating networks for a diverse variety of design scenarios, including: urban street layouts~(a-c), floorplanning for large facilities such as offices and hospitals~(d), and game level designs~(e). The user simply specifies an input mesh presenting the problem domain, of which the edges present potential placements of network segments, and some high-level specifications of the functions of the generated network. Examples include a preference for interior-to-boundary traffic~(a) or interior-to-interior traffic~(b), networks with specific end points (i.e., {\em sinks}) on the boundary~(c,d) or sinks within~(e), and local feature control such as reducing T-junctions~(b) and forbidding dead-ends~(c).}
   \label{fig:teaser}
 }

\maketitle

\begin{abstract}

Connectivity and layout of underlying networks largely determine the behavior of many environments. For example, transportation networks determine the flow of traffic in cities, or maps determine the difficulty and flow in games. Designing such networks from scratch is challenging as even local network changes can have large global effects. We investigate how to computationally create networks starting from {\em only} high-level functional specifications. Such specifications can be in the form of network density, travel time versus network length, traffic type, destination locations, etc. We propose an integer programming-based approach that guarantees that the resultant networks are valid by fulfilling all specified hard constraints, and score favorably in terms of the objective function. We evaluate our algorithm in three different design settings (i.e., street layout, floorplanning, and game level design) and demonstrate, for the first time, that diverse networks can emerge purely from high-level functional specifications.

\end{abstract}

\keywordlist

\TOGlinkslist


\section{Introduction}

Layout computation is an important tool for modeling virtual environments and planning new environments for construction. The behavior of such environments is largely dictated by the underlying networks, both at global and local scales. For example, in the context of urban planning, the transportation network determines the access patterns across a city; or, in the context of games, a well-designed layout map can ensure graded complexity of the play area.

Often, a designer would want to create networks simply by describing how the target environment should behave. We refer to such high-level descriptions as {\em functional specifications}. For example, functional specifications can come in the form of desired network density, transportation pattern, local features (e.g., whether dead-ends or branches are allowed), etc.
In this paper, we study how to design networks starting {\em only} from such functional specifications.

Limited support exists for such a design paradigm. A brute force solution is to propose various network variations and rank them in an effort to design a target environment. For ranking, one can evaluate the behavior of a given network using a black-box forward simulation (e.g., using a traffic simulator, or solving a flow problem).
However, such a trial-and-error approach is tedious and time consuming.
This is especially so as even small changes in network topology can result in large global behavioral changes.

We observe that in network design one has to typically balance between conflicting requirements: networks should be densely connected in order to obtain low average transportation times; while, the total length of the network should be small in order to leave space for other assets.
In addition, there are certain factors that may have large effects on the appearances of networks: local features, such as dead-ends and branches, and the distribution of the transportation destinations (denoted as {\em sinks} in this paper). 
%
Based on these observations, we produce a desirable network layout based on a novel integer programming (IP)-based approach that takes as input a set of functional specifications and boundary description of the target domain.
Technically, the proposed IP formulation guarantees that the designed networks are {\em valid} by ensuring that they are free of islands (see Figure~\ref{fig:IP_valid} for examples) and offer sufficient coverage over the target domain, while having desirable {\em quality} as measured by the specified functional specifications.


We evaluate the proposed algorithm in the context of urban street layouts, floorplanning, and game level design. We identify a set of commonly appearing functional specifications (Section~\ref{sec:functional}) across the three scenarios and propose how to effectively model them (Section~\ref{sec:IP}). 
For example, Figure~\ref{fig:teaser} shows different networks created by our algorithm using different functional specifications.

In summary, our main contributions are:
\begin{itemize}
\item proposing the problem of network design directly from functional specifications;
\item formulating the problem as an integer optimization framework that supports common functional specifications; and
\item evaluating the generated networks in three different design contexts, and demonstrating that non-trivial and desirable network layouts can emerge only from high-level functional specifications.
\end{itemize}

\section{Related Work}

\mypara{Layout modeling}
There are various layouts that have been modeled in computer graphics. In some of these layout computations, transportation networks have no significant role, e.g., in the distribution of vegetation~\cite{Deussen:1998:RMR}. However, there are many examples of layouts that have at least some network aspect to them. In furniture layouts of Yu et al.~\shortcite{Yu:2011:MIH}, the existence of obstacle-free walking paths was a consideration in modeling the objective function. Several biologically inspired simulations also model networks. For example, Runions et al.~\shortcite{Runions:2005:MAV} generate leaf venation patterns that result in fascinating graphs. River networks can be modeled using hydrological principles~\cite{Genevaux:2013:TGU}.
Another important problem is the layout of building floorplans~\cite{Merrell:2010:CGR} or game levels~\cite{Ma:2014:GLL}, because the rooms and hallways also induce a network.


\mypara{Street modeling}
Initial work on street network modeling focused on algorithms to synthesize street networks that resemble existing ones. One approach is to grow street segments greedily until the available space is filled~\cite{Parish:2001:PMC,Weber:2009:IGS}. An alternative version is to first sample points on the street network that are connected in a subsequent algorithm step~\cite{Aliaga:2008:IEB}. Chen et al.~\shortcite{Chen:2008:IPS} proposed the use of tensor fields to guide the placement of street segments.
One way to improve synthesis algorithms is to optimize the quality of street networks to include local geometric and functional quality metrics, such as street network descriptors~\cite{AYWN2014}, sunlight for resulting buildings~\cite{Vanegas:2012:IDO}, the shape of individual parcels~\cite{Yang:2013:UPL,Peng:2014:CLD:2601097.2601164}, or the shape of individual roads interacting with the environment~\cite{Marechal:2010:PGO}.
There are some initial attempts to include global traffic considerations into the layout process. A simple first step is to compute a traffic demand model and use this model to modify street width or to guide expansion of the street network~\cite{Weber:2009:IGS,Vanegas:2009:IDO}. The connectivity of the road network is also a fundamental requirement for generating high-level roads connecting cities and villages~\cite{Galin:2012:AHR}. A recent paper describes how to design traffic behavior in an urban environment~\cite{Garcia:DLS:2014}. This paper touched on many aspects of traffic design that would make a great addition to our proposed system. While most of the proposed components are complementary to our paper, one important component of this system is an algorithm to modify an existing street network by making low-level random modifications. 
Instead, we focus on the complementary problem of generating the initial coarse network only from functional specifications.

\mypara{Modeling methodology}
From the methodology side, there have been multiple interesting approaches in the recent literature that are suitable to model interesting shapes.
The first approach is procedural modeling using grammars, e.g.,~\cite{Prusinkiewicz:1990:ABP}, that enables a user to write rules for shape replacement.
The second approach is to learn new shapes when given a database of existing shapes, e.g.,~\cite{Kalogerakis:2012:ShapeSynthesis}. This approach requires machine learning techniques, such as graphical models, to encode the relationships between different shape components. The third approach is to use optimization to compute forms from a functional description, e.g.,~\cite{bymw_goodLayout_sigg13}. Our methodology follows the last line of work.

Network design by IP has also been explored in other fields (e.g., telecommunication~\cite{5534462}, transportation~\cite{Luathep2011808}). However, the approaches are quite different -- we consider networks as graphs to be embedded in a 2D plane, while other approaches often consider networks as abstraction models.

\section{Functional Specifications}
\label{sec:functional}

Our goal is to allow users to create networks simply by describing a set of {\em functional specifications} on how the generated networks should behave.
We studied commonly used design practices from the relevant literature~\cite{Southworth:1995:SSA,Meyer:00,planStreet:03,Southworth03,Marshall:2011:SP,TRB:2010:HCM}. From these works, we recommend~\cite{APA:06} for a simple introduction to urban planning standards in general and \cite{Ortuzar:11} for a more technical introduction to transportation models. In the context of floorplanning, circulation is a fundamental concept in architecture (cf.,~\cite{Ching:1996:AFSO}) that is considered in the design of every building.
For games, we referred to RPG game maker~\cite{RPG} and different game design blogs (e.g., http://www.vg-leveldesign.com/level-layout-types/ and online book http://pcgbook.com/). 
We identified a set of recurring considerations and summarize them as functional specifications that we describe next.

\mypara{Density} The desired density of a network encodes the average spacing between the network edges.

\mypara{Network lengths versus travel distances} For networks with comparable densities, two extreme cases can happen. On one extreme,  the total network length can be minimized. On the other extreme, networks can facilitate more efficient travels, usually toward certain destination locations predefined on the domain. In our framework, the user can give a relative preference between the two.

\mypara{Traffic types} We support three types of traffic for a target network: interior-to-boundary, interior-to-interior, and boundary-to-boundary traffic. Networks arising from different types of traffic specifications tend to look quite different (e.g., Figure~\ref{fig:results_plainview}a, e, d, in respective order). Users can indicate a preference among the three.

\mypara{Sink locations} 
A key function of networks is to facilitate accesses to certain predefined destination locations, i.e., sinks. We allow the shape of a network to be controlled by the distributions of such sinks. Intuitively, a network tends to look like a root-like structure grown from the destination locations. The user can select the sink location to be: (i)~all of the boundary, or (ii)~only at a subset of the boundary, or (iii)~at the interior of the target domain.

\mypara{Local features} There are certain local features that can have profound effects on the appearances of the target networks. Examples are dead-ends, branches, and T-junctions. Users can determine if any of such features should be forbidden or just appear with lower frequency.

\mypara{User specifications} We provide two ways to directly control the generated networks. First, specifying certain locations as {\em obstacles} that the generated network must avoid. Second, enforcing certain routes to appear in the generated network, for example, a direct route between two boundary locations to boost through-traffic.

\section{IP-based Network Design}
\label{sec:IP}

In this section, we introduce an integer-programming (IP)-based optimization approach for designing networks from the functional specifications described in Section~\ref{sec:functional} on a given problem domain. We assume that the domain, given as a piecewise linear 2D polygon, is discretized into a polygonal mesh, $M$. Our goal is to select a subset of the edges in $M$ as the network. We consider a selected network to be \emph{valid} if it satisfies two constraints:
(i)~A {\em no-island constraint} that ensures accesses to specified destination locations (i.e., {\em sinks}) predefined on the domain.
(ii)~A {\em coverage constraint} that ensures the network sufficiently covers the whole domain (i.e., all vertices in $M$ are within a distance to the network).

We rate such valid networks based on a set of \emph{quality measures}: First, we prefer networks with smaller total length as longer networks are often more expensive to build and maintain, and take up more usable space from the domain. Second, we prefer networks with shorter travel distances to the sinks (from every vertex in the network). As these are mutually competing goals, our objective function measures the quality of the solutions as a weighted sum of the two.

We also support two additional quality measures: (i)~ensure quick accesses between any two locations in the network by a {\em point-to-point constraint} (note that this is not guaranteed if we only optimize for quick accesses to the sinks.); and (ii)~introduce measures for controlling the local features (e.g., dead-ends, branches, and T-junctions) in a network. Before describing how we encode such quality measures, we begin with the following definitions.

\begin{figure*}[tbh!]
\centering
\includegraphics[width=\linewidth]{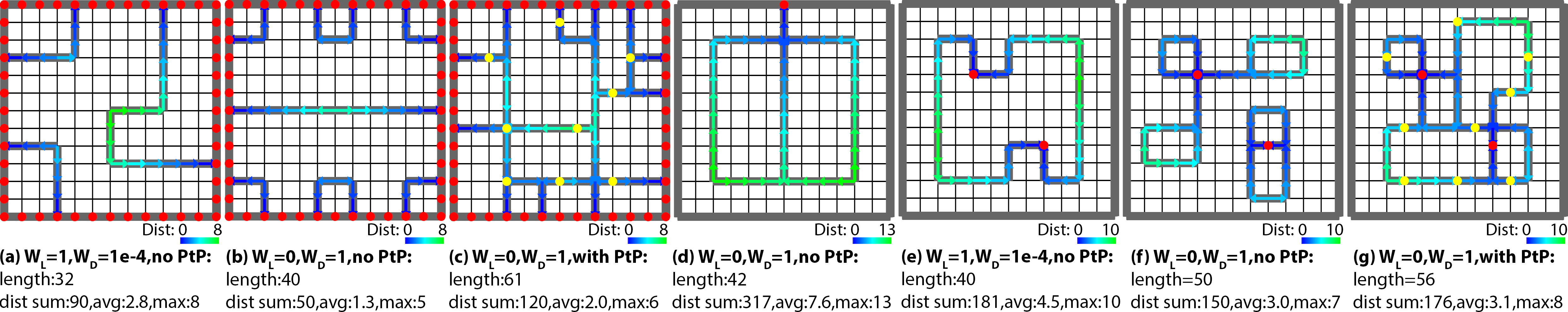}
\caption{Example results. Sink vertices are marked in red. The coverage range is two edges wide. The distance values of active half-edges are colored (see legends for color ranges). Boundary edges are excluded from the calculations. We forbid dead-ends and edges that are too close to each other. (a) to (c): A typical urban layout scenario such that all boundary vertices are sinks. The first two use different weights to optimize for (a) numbers of network edges and (b) sum of distance values. Note that we do not specifically constrain the maximum of distance values. (c): A result that also optimizes for distance values but with the point-to-point constraint enabled (sampled vertices are marked in yellow). (d): We now constrain a boundary vertex (top middle) to be the sole sink. (e) to (g): A typical floorplan scenario such that a few inner vertices are sinks (e.g., elevators). Similarly, they differ by different optimization weights and whether point-to-point constraint is enabled.}
\label{fig:IP_results}
\end{figure*}

\begin{definition}
\label{def:network}
A {\em network} is a subset of the edges in $M$. An edge is {\em active} if it is in the subset; otherwise, it is {\em inactive}. A vertex is {\em active} if any of its adjacent edges are active; otherwise, it is {\em inactive}. The {\em sinks} are a predefined subset of the vertices in $M$. 
\end{definition}

We use Boolean indicator variable, $E_m$, to model the active/inactive states of every edge, $e_m$, for $m \in [0, N_E)$ with $N_E$ being the number of edges in $M$.

\begin{figure}[tb!]
\centering
\includegraphics[width=\linewidth]{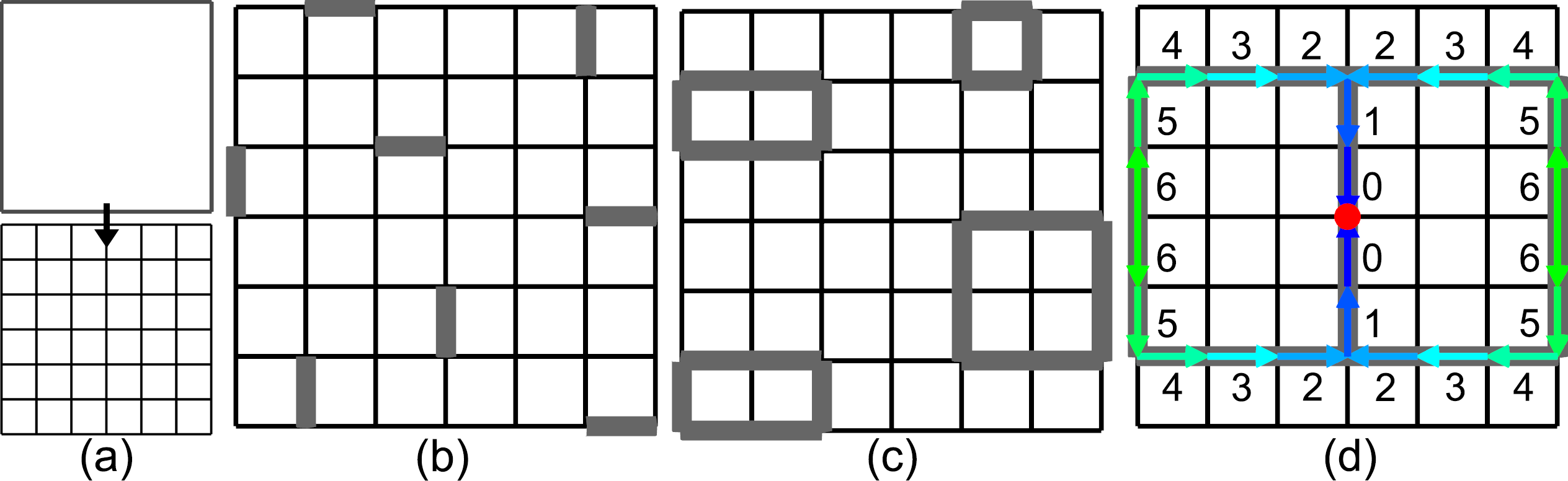}
\caption{The impact of no-island constraints. The coverage range is one edge wide. (a)~The problem domain is first discretized into a mesh. (b)~A network (i.e., a subset of the edges) that sufficiently covers the domain using the fewest possible number of edges. However, the solution may consist of many disconnected parts. (c)~Simply forbidding degree-1 vertices is not enough to guarantee the network to be free of islands, as it is still possible to form disconnected loops. (d)~Our no-island constraint guarantees that the network is entirely connected to the sink (red vertex). Here we show the distance values assigned to each active half-edge.}
\label{fig:IP_valid}
\end{figure}

\mypara{No-island constraint} Our goal is to design networks {\em without} islands, while still allowing loops in the networks. We proceed as follows: First, without changing the definition of a network, we distinguish each half-edge, $e_{i\rightarrow j}$ (i.e., goes form vertex $v_i$ to $v_j$), as active or inactive, by a Boolean indicator variable $E_{i\rightarrow j}$, for $i,j \in [0, N_V)$ with $N_V$ being the number of vertices in $M$. Our goal is to assign each active half-edge, $e_{i\rightarrow j}$, a {\em distance value} that roughly encodes the distance of $v_j$ toward the closest sink vertex along the active half-edges in the network. We model the distance value of half-edge $e_{i\rightarrow j}$ as a non-negative continuous variable, $D_{i\rightarrow j}$, for $D_{i\rightarrow j} \in [0,\Delta_{all}] $ with $\Delta_{all}$ being the sum of the lengths of all half-edges in $M$.
To achieve this goal, the following requirement should be satisfied:
\begin{proposition}
\label{equ:IP_proposition}
For a half-edge $e_{i\rightarrow j}$ to be active, at least one of its succeeding half-edges (i.e., the half-edges that go from $v_j$ but not go back to $v_i$) also needs to be active and be assigned a distance value smaller than the distance value of $e_{i\rightarrow j}$, except if $v_j$ is a sink vertex.
\end{proposition}

We model this requirement as:
For every succeeding half-edge, $e_{j\rightarrow k}$, for $k \in \mathcal{N}_j \setminus \{i\}$, where $\mathcal{N}_j$ is the one-ring neighborhood of $v_j$, of every half-edge $e_{i\rightarrow j}$ in $M$ such that $v_j$ is not a sink vertex,
\begin{equation}
L_{i\rightarrow j;j\rightarrow k} \le E_{j\rightarrow k} \;\; \text{and}
\label{equ:IP_validity0}
\end{equation}
\begin{equation}
D_{j\rightarrow k} - D_{i\rightarrow j} + \Delta_{all} L_{i\rightarrow j;j\rightarrow k} \le \Delta_{all} - \Delta_{i\rightarrow j},
\label{equ:IP_validity1}
\end{equation}
where, $L_{i\rightarrow j;j\rightarrow k}$ are auxiliary Boolean variables associated with every half-edge, $e_{i\rightarrow j}$, one for each of its succeeding half-edges, $e_{j\rightarrow k}$, for $i,j \in [0,N_V)$, $k \in [0, K)$. $\Delta_{i\rightarrow j}$ is the length of half-edge $e_{i\rightarrow j}$. Note that $L_{i\rightarrow j;j\rightarrow k}$ can be true only if: (i)~$e_{j\rightarrow k}$ is active (enforced by inequality~\ref{equ:IP_validity0}), and (ii)~$e_{j\rightarrow k}$ has a smaller distance value than the distance value of $e_{i\rightarrow j}$ by at least the length of $e_{i\rightarrow j}$ (enforced by inequality~\ref{equ:IP_validity1}). 

Next, the following inequality ensures that for $e_{i\rightarrow j}$ to be active, at least one of its auxiliary variables must be true:
For every half-edge in $M$, $e_{i\rightarrow j}$, such that $v_j$ is not a sink vertex,
\begin{equation}
E_{i\rightarrow j} - \sum_{0 \le k < K} L_{i\rightarrow j;j\rightarrow k}  \le 0.
\label{equ:IP_validity2}
\end{equation}

\begin{wrapfigure}{r}{0.15\columnwidth}
\hspace{-20pt}
   \centering
     \includegraphics[width=0.17\columnwidth]{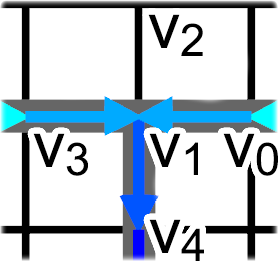}
\end{wrapfigure}
To illustrate no-island constraints, on the right, we show the top-middle part of Figure~\ref{fig:IP_valid}(d) with vertex indices. As all edges have uniform lengths, constants $\Delta_{i\rightarrow j}$, $i, j \in [0,N_V)$, all equal to one. Continuous variables $D_{0\rightarrow 1}$, $D_{1\rightarrow 2}$, $D_{1\rightarrow 3}$, and $D_{1\rightarrow 4}$ are assigned to be $2$, $0$ (arbitrary), $2$, and $1$. Therefore, by inequality~\ref{equ:IP_validity1}, $L_{0\rightarrow 1;1\rightarrow 2}$ and $L_{0\rightarrow 1;1\rightarrow 3}$ are false and $L_{0\rightarrow 1;1\rightarrow 4}$ is true. This allows $E_{0\rightarrow 1}$ to be true by inequality~\ref{equ:IP_validity2}.

We now prove that that the requirement in Proposition~\ref{equ:IP_proposition} forbids islands in networks. Assume a network contains one or more islands and the requirement is still met. There exists at least one weakly connected component that is not connected to the sink vertices (i.e., an island). Within this island, there cannot exist any active half-edge that has no succeeding active half-edges, otherwise the requirement is immediately violated. Therefore, there must exist at least one loop of active half-edges in the island, denoted as $e_{0\rightarrow 1}$, $e_{1\rightarrow 2}$, ... $e_{n-1\rightarrow 0}$, $n$ is the number of vertices in the loop. Since none of these vertices are sinks, $D_{0\rightarrow 1} > D_{1\rightarrow 2} > ...  > D_{n-1\rightarrow 0} > D_{0\rightarrow 1}$, a contradiction.

Note that the requirement does not forbid loops in networks. One example is shown in Figure~\ref{fig:IP_valid}.

Finally, we say that an edge is active if and only if at least one of its two half-edges is active:
For every edge in $M$, $e_x$,
\begin{equation}
-1 \le E_{i\rightarrow j} + E_{j\rightarrow i} - 2 E_x \le 0,
\label{equ:IP_coverage0}
\end{equation}
where, $E_{i\rightarrow j}$ and $E_{j\rightarrow i}$ are the Boolean indicator variables of $e_x$'s two half-edges. $E_x$ is the Boolean indicator variable of $e_x$.

The above formulation above was inspired by an IP formulation of the Traveling Salesman Problem~\cite{Miller:1960:IPF:321043.321046}. However, unlike theirs, the new formulation allows loops involving the starting nodes (i.e., sink vertices).

\mypara{Coverage constraint} We expect a network to sufficiently cover the whole domain. We take a simple coverage model such that an active vertex covers itself and its nearby vertices within a distance threshold. Alternatively, different coverage models (to determine which vertices are covered by an active vertex) can be used for different design scenarios. A network sufficiently covers $M$ if all the vertices in $M$ are covered. We model this as described next.

We denote the active/inactive states of every vertex in $M$, $v_y$, as Boolean indicator variables $V_y$, $y \in [0,N_V)$. Since a vertex is active if and only if at least one of its adjacent edges is active, we have the following constraint:
For every vertex in $M$, $v_y$,
\begin{equation}
1 - |\mathcal{E}_y| \le \sum_{0 \le x < X_0} E_x - |\mathcal{E}_y| V_y \le 0,
\label{equ:IP_coverage1}
\end{equation}
$x \in \mathcal{E}_y$, where $\mathcal{E}_y$ is the set of edges that are adjacent to $v_y$.

We can now encode the coverage requirement as:
For every vertex in $M$, $v$,
\begin{equation}
\sum_x V^{cover}_{x} \ge 1,
\label{equ:IP_coverage2}
\end{equation}
where $V^{cover}_{x}$ denotes the set of indicator variables of the vertices that cover $v$.

\begin{figure*}[tbh!]
\centering
\includegraphics[width=\linewidth]{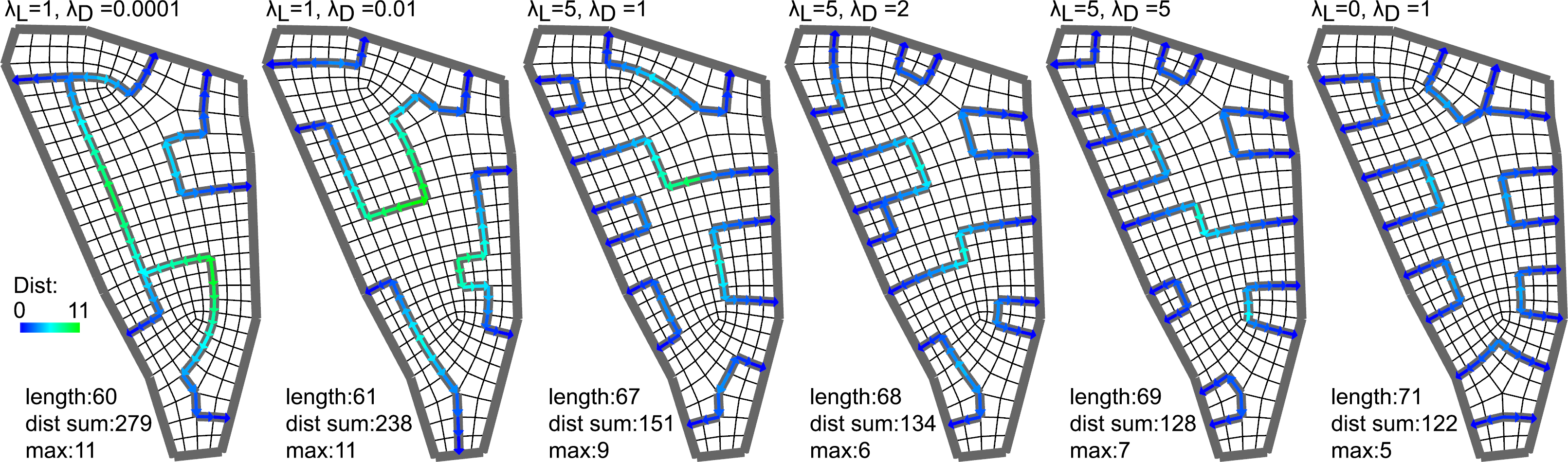}
\caption{Optimization results with a gradual change of $\lambda_L$ (to minimize network lengths) versus $\lambda_D$ (to minimize travel distances). In general, a larger $\lambda_L$ leads to smaller network lengths but larger travel distances, while a larger $\lambda_D$ leads to the opposite. We use the same setting as in Figure~\ref{fig:IP_results}.}
\label{fig:IP_gradual}
\end{figure*}

\mypara{Objective function} We want to minimize two aspects of a network. First, the total length of the network. Second, the total travel distances to the sinks. The first term can be expressed as the summation of all edge indicator variables multiplied by each edge's length. The second term can be expressed as the summation of all distance values of half-edges. Thus the objective function takes the form:
\begin{equation}
\min_{E_{i\rightarrow j}, D_{i\rightarrow j}, L_{i\rightarrow j;j\rightarrow k}, E_x, V_y} \quad \lambda_L \sum_{x} \Delta_x E_x + \lambda_D \sum_{i,j} D_{i\rightarrow j},
\label{equ:IP_obj0}
\end{equation}
where $\Delta_x$ is the length of edge $e_x$, $\lambda_L$ is the weight for the total length term, and $\lambda_D$ is the weight for the total travel distance term. Note that when $\lambda_D$ is set to zero, the distance values may not be assigned correctly and have to be calculated in a post-process. In Figure~\ref{fig:IP_gradual}, we analyze how the assignments of $\lambda_L$ versus $\lambda_D$ affect the optimization results.

\mypara{Point-to-point constraint} Here, our goal is to constrain the travel distances between any two vertices in the network, not just the travel distances to the sinks. While explicitly modeling such constraints is possible, it would be prohibitively expensive since we need to model every possible paths between every possible pairs of vertices. Below, inspired by the construction of $k$-spanners for graphs~\cite{Baswana:2007:SLT:1255378.1255381}, we describe a cost-effective way to approximate our goal.


We first partition $M$ into a set of sub-meshes (i.e., a connected set of faces), $M_0, M_1,... , M_{N_S-1}$, $N_S$ is the number of sub-meshes. $M_i \cap M_j = \emptyset$, $i \ne j$, $0 \le i,j < N_S$. $M_0 \cup M_1 ... \cup M_{N_S-1} = M$. We assume the partition is given by the user. We say that two sub-meshes are adjacent to each other if they share common edges. Next, we randomly sample one vertex in every sub-mesh. For sampling, we consider vertices that are not adjacent to any other sub-meshes, unless such vertices do not exist. For every pair of adjacent sub-meshes, we red exhaustively enumerate the set of paths (i.e., a consecutive sequence of edges) connecting the two sampled vertices with topological lengths not greater than the length of a shortest path between the two vertices plus a tolerance value (we use $2$).
\begin{wrapfigure}{r}{0.24\columnwidth}
\hspace{-20pt}
   \centering
     \includegraphics[width=0.28\columnwidth]{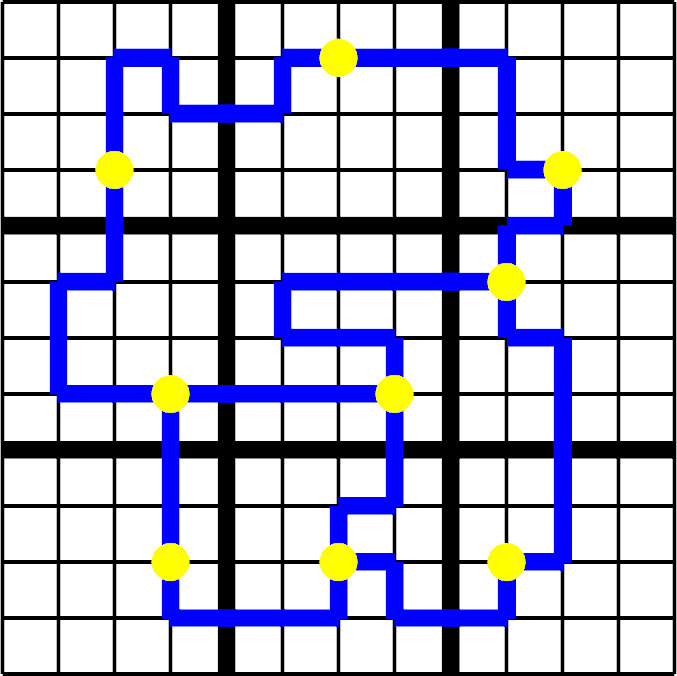}
   \vspace{-10pt}
\end{wrapfigure}
For every such set, we require that at least one of the paths is active  (i.e., consisting of active edges). In summary, we require the network to connect every pair of adjacent sub-meshes by connecting their respective sampled vertices. One example (for Figure~\ref{fig:IP_results}c and g) is shown on the right. They are modeled as follows.

Let $P_{a\rightarrow b,x}$ be a Boolean indicator variable indicating the presence of the $x$-th path among the set of paths connecting two sampled vertices $v_a$ (of sub-mesh $M_a$) and $v_b$ (of sub-mesh $M_b$), for $a,b \in [0,N_S)$ and $x \in [0,X_2)$ with $X_2$ being the size of the set. Let $E^{a\rightarrow b,x}_n$, $n = 0, 1, ..., N-1$ denote the edges on path $P_{a\rightarrow b,x}$, $N$ is the size of the path. We have:
\begin{equation}
-N+1 \le N P_{a\rightarrow b,x} - \sum_{n} E^{a\rightarrow b,x}_n \le 0.
\label{equ:IP_PtP0}
\end{equation}
For every set of paths connecting sampled vertices $v_a$ and $v_b$:
\begin{equation}
\sum_{x} P_{a\rightarrow b,x} \ge 1.
\label{equ:IP_PtP1}
\end{equation}

As shown in Figure~\ref{fig:IP_results}c and g, enforcing such constraints leads to networks that are more tightly connected, which in turn have quicker accesses between any two vertices in the network. Users can control the strength of the point-to-point constraint by the density of the partitions. While this approach is computationally cheap, it can overconstrain the resulting network. Therefore, we consider alternative ways that better balance flexibility and computational cost as a direction for future work.

Next, we introduce quality measures for controlling local features in a network.

\mypara{Dead-end avoidance} We may desire networks that have no dead-ends, i.e., an active vertex that is adjacent to just one active edge. To achieve this, we give every half-edge, $e_{i\rightarrow j}$ (from vertex $v_i$ to $v_j$), a {\em non-emptiness} Boolean indicator variable, $\nu_{i\rightarrow j}$. $\nu_{i\rightarrow j}$ is true if any of the $v_j$'s adjacent edges, excluding the edge of $e_{i\rightarrow j}$, is active. It is false otherwise. It is modeled as follows.

For every half-edge in $M$, $e_{i\rightarrow j}$,
\begin{equation}
-|\mathcal{E}_j \setminus \{i\rightarrow j\}|+1 \le \sum E_x - |\mathcal{E}_j \setminus \{i\rightarrow j\}| \nu_{i\rightarrow j} \le 0,
\label{equ:IP_deadend0}
\end{equation}
$x \in \mathcal{E}_j \setminus \{i\rightarrow j\}$, where $\mathcal{E}_j$ is the set of edges adjacent to $v_j$.

Dead-ends can then be avoided by the following constraints:

For every half-edge in $M$, $e_{i\rightarrow j}$,
\begin{equation}
E_{i\rightarrow j} -  \nu_{i\rightarrow j} \le 0.
\label{equ:IP_deadend1}
\end{equation}

\mypara{Branch avoidance} It is simple to avoid branches (i.e., a vertex with more than three adjacent active edges) in a network by the following constraint:
For every vertex in $M$, $v_y$,
\begin{equation}
\sum E_x \le 2,
\label{equ:IP_branch0}
\end{equation}
$x \in \mathcal{E}_y$, $\mathcal{E}_y$ is the set of edges adjacent to $v_y$.
Enabling branch avoidance constraint leads to cycle-like networks (see the game level design example in Figure~\ref{fig:results_gamelevel}b).

\mypara{Local configuration control} We identify certain local
\begin{wrapfigure}{r}{0.24\columnwidth}
\hspace{-20pt}
   \centering
     \includegraphics[width=0.28\columnwidth]{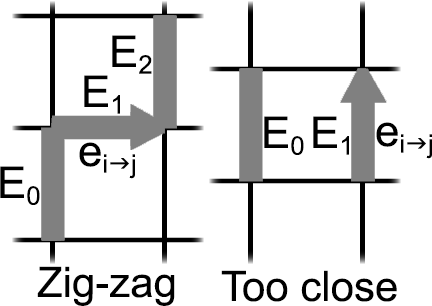}
   \vspace{-10pt}
\end{wrapfigure}
configurations of active edges as undesirable, which include: (i)~zig-zags and (ii)~edges that are too close to each other. Their occurrences can be strictly forbidden or minimized.

\begin{figure*}[t!]
\centering
\includegraphics[width=\linewidth]{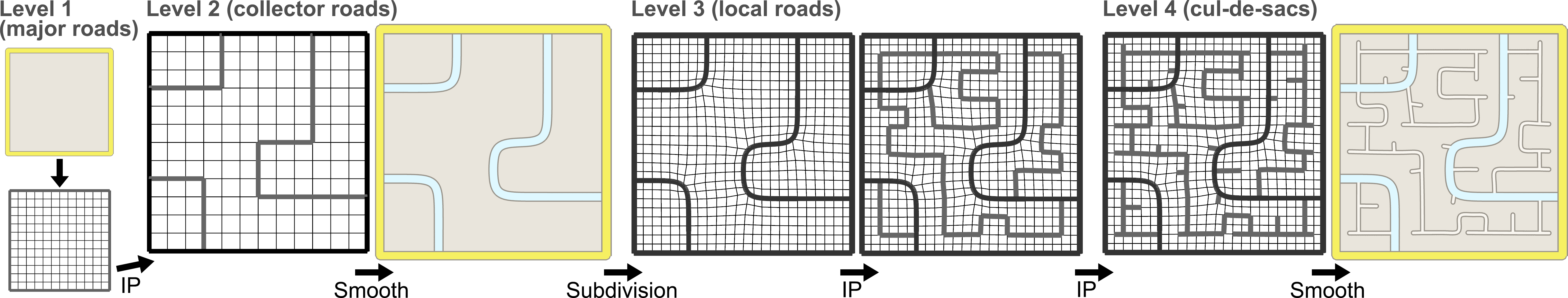}
\caption{Generating street layouts. The first level layout (major roads) is designed by the user. Afterwards, for each sub-region, we generate layouts in three levels of decreasing coverage ranges. For each level, a rough street network is first generated by the IP-based approach (shown in gray). Afterwards, the geometry of the generated street network is realized by a smoothing process (the last two levels are smoothed in one pass). The mesh is subdivided at the third level for increased degrees-of-freedom. Dead-ends are typically allowed only at the last level.}
\label{fig:results_pipeline}
\end{figure*}

As shown on the right, for each half-edge, $e_{i\rightarrow j}$, we identify two undesirable configurations. The presence of each configuration on $e_{i\rightarrow j}$ is denoted by a Boolean indicator variable, $Z^k_{i\rightarrow j}$, where $k$ equals $0$ for zig-zags and $1$ for edges that are too close to each other. This is modeled as follows: For every $Z^k_{i\rightarrow j}$ that denotes the presence of the $k$-th undesirable configuration on half-edge $e_{i\rightarrow j}$,
\begin{equation}
0 \le \sum E_x - |E_x| Z^k_{i\rightarrow j} \le |E_x|-1,
\label{equ:IP_undesirable0}
\end{equation}
$x \in \mathcal{E}_{i\rightarrow j, k}$, where $\mathcal{E}_{i\rightarrow j, k}$ denotes the set of edges comprising the $k$-th undesirable configuration on $e_{i\rightarrow j}$.

To forbid the presence of any of such configurations, simply enforce all $Z^k_{i\rightarrow j}$ to be false. As this constraint can be too strict, we can instead minimize the occurrence of such configurations by adding the weighted summation of $Z^k_{i\rightarrow j}$ to the objective function.

In a similar way, we can forbid or minimize the occurrence of T-junctions as follows. For each half-edge pointing to a valence-$4$ vertex, $e_{i\rightarrow j}$, the presence of a T-junction on $e_{i\rightarrow j}$ is denoted by a Boolean indicator variable, $T_{i\rightarrow j}$, modeled as follows:

For every half-edge in $M$, $e_{i\rightarrow j}$,
\begin{equation}
0 \le E_1 + E_2 + E_3 + (1-E_0) - 4 T_{i\rightarrow j} \le 3,
\label{equ:IP_undesirable1}
\end{equation}
\begin{wrapfigure}{r}{0.16\columnwidth}
\hspace{-16pt}
   \centering
     \includegraphics[width=0.16\columnwidth]{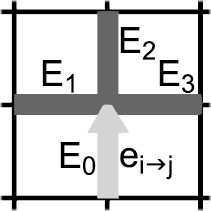}
   \vspace{-10pt}
\end{wrapfigure}
where, $E_0$ is the edge indicator variable of $e_{i\rightarrow j}$'s edge, and $E_1$ to $E_3$ are the edge indicator variables of the other three edges adjacent to $v_j$ (see right figure). Again, we can forbid T-junctions by enforcing all $T_{i\rightarrow j}$ to be false, or minimize their occurrence by adding the weighted summation of $T_{i\rightarrow j}$ to the objective function.

We now rewrite the objective function as follows:
\begin{equation}
\begin{split}
\min_{e_{i\rightarrow j}, D_{i\rightarrow j}, L_{i\rightarrow j;j\rightarrow k}, E_x, V_y, Z^k_{i\rightarrow j}, T_{i\rightarrow j}} \quad \lambda_L \sum_{x} \Delta_x E_x +\\
\lambda_D \sum_{i,j} D_{i\rightarrow j} + \sum_{k,i,j} \lambda^k_{Z} Z^k_{i\rightarrow j} + \lambda_T \sum_{i,j} T_{i,j},
\label{equ:IP_obj1}
\end{split}
\end{equation}
where, $\Delta_x$ is the length of edge $E_x$, $\lambda_L$ is the weight for the total length term, $\lambda_D$ is the weight for the total travel distance term, $\lambda^k_{Z}$, $k=0, 1$, are the weights for minimizing the occurrences of the two undesirable configurations, and $\lambda_T$ is the weight for minimizing the occurrence of T-junctions.

\mypara{User specifications} Users can explicitly specify certain combinations of vertices and/or edges to be inactive or active in a network. It is straightforward to impose these specifications by constraining the corresponding vertex or edge indicator variables to be true or false. Examples of such specifications can be seen in Section~\ref{sec:result}.

We show example results in Figure~\ref{fig:IP_results}. In summary, the IP formulation consists of a linear objective function in Equation~\ref{equ:IP_obj1} and linear constraints in Equations~\ref{equ:IP_validity0} - \ref{equ:IP_deadend1},~\ref{equ:IP_undesirable0}, and~\ref{equ:IP_undesirable1}.

\section{Applications and Results}
\label{sec:result}

Our results are categorized by different design scenarios: urban street layouts (Section~\ref{sec:urban}), floor plans for large facilities such as offices and hospitals (Section~\ref{sec:floorplan}), and game level design (Section~\ref{sec:game}). The large variety of the results demonstrate the versatility of our IP-based approach.

\subsection{Urban street layouts}
\label{sec:urban}

We aim at generating street layouts at the scale of city blocks to a small city. Based on the hierarchical nature of real-world road networks, we lay out the streets in four levels. First, a coarse network of {\em major roads} (e.g., freeways or arterial roads) that partitions the city into several sub-regions. Second, for each sub-region, a denser network of {\em collector roads} with the purpose of collecting traffic from the local roads. Third, grown from the collector roads, an even denser network of {\em local roads} that roughly span the whole sub-region. Fourth, there may be some {\em cul-de-sacs} grown from the streets at the previous two levels.

We assume that the network of major roads is designed by the user. Afterwards, for each sub-region, our pipeline to create the street layouts is as follows (see Figure~\ref{fig:results_pipeline}).

\begin{enumerate}
\item We first compute a dense network of street segment candidates in the form of a semi-regular (i.e., most vertices are of valence $4$) quad mesh, $M$, wherein the quads are roughly of uniform size and their shapes are close to a square. This assumption is based on the observation that real-world urban street layouts often favor 90-degree intersections. In practice, the input problem domain is quadrangulated by the patch-wise quadrangulation algorithm in Peng et al.~\shortcite{Peng:2014:EQ:2577382.2541533}.
\item Beginning at the sub-region level, an initial street network is computed by selecting a subset of segments of the dense network using the IP-based approach described in Section~\ref{sec:IP}.
\item The geometry of the street network (i.e., positions of the vertices along the street edges) is further improved by a snake-based smoothing algorithm described in the appendix.
\item If the last level is not reached, we generate a denser network for the next lower level. To do so, we need to increase the resolution of the mesh by a Catmull-Clark subdivision scheme without vertex repositioning for greater degrees of freedom of the IP computation. The process starting with step 2 is then repeated at a lower level on the subdivided mesh.
\end{enumerate}

\mypara{Functional specifications} A user can specify the function of the street network using the terms discussed in Section~\ref{sec:IP} as follows.

\begin{figure}[b!]
\centering
\includegraphics[width=\linewidth]{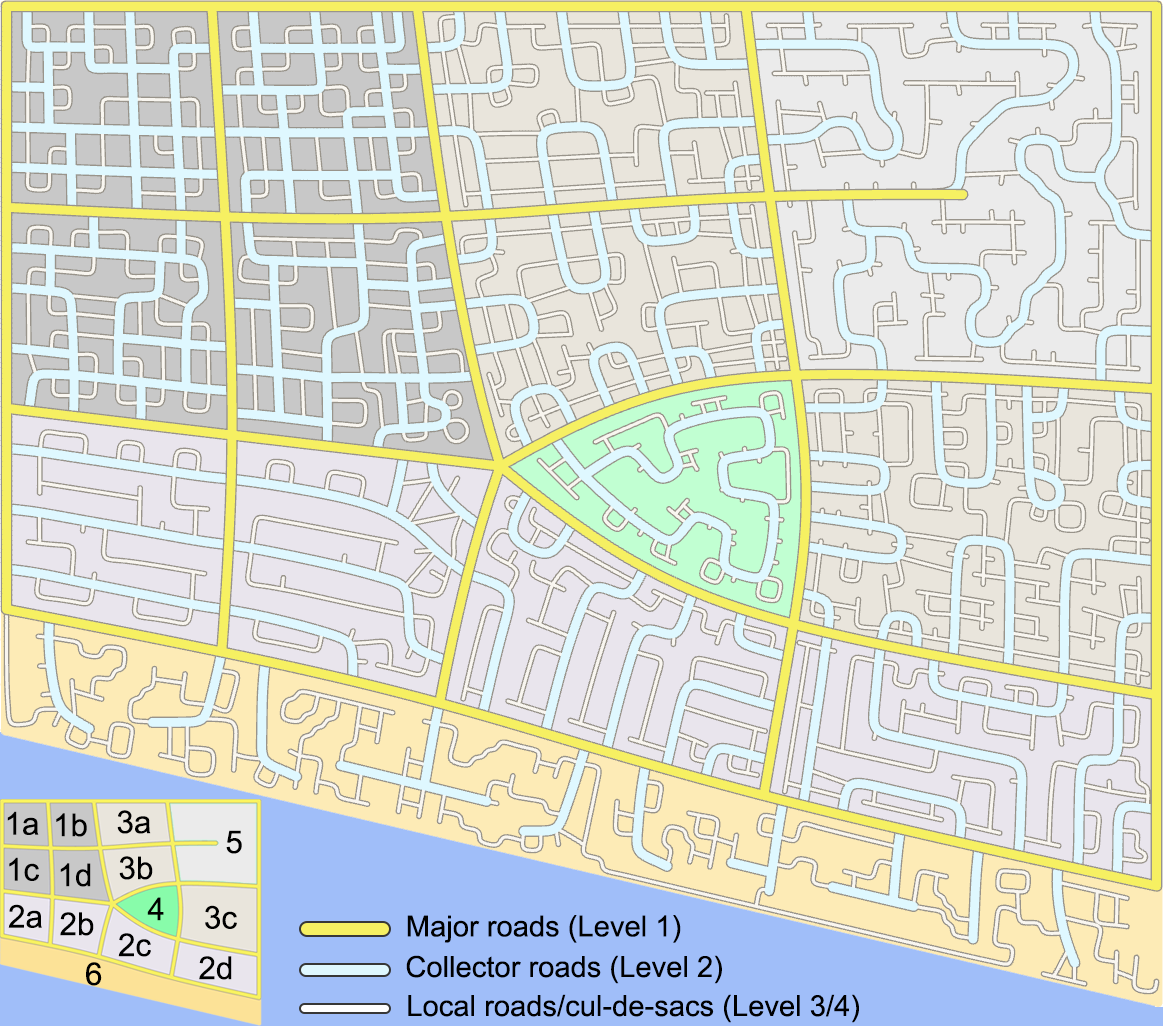}
\caption{A city-level street layout. To mimic a coastal town setting, different functional specifications are used for different sub-regions: (1a-d) Downtown layouts with a higher density, a stronger interior-to-interior traffic, and no dead-ends. (2a-d) and (3a-c) Suburban layouts that either favor (2a-d) shorter network lengths or (3a-c) shorter travel distances to the boundaries. (4) A gated community with a single entrance to the boundary. (5) A lower density part of the city with a sparser layout. (6) A beach-front area. Dead-ends are specifically allowed for the level-2 roads.}
\label{fig:results_seaside}
\end{figure}

\begin{figure*}[t!]
\centering
\includegraphics[width=\linewidth]{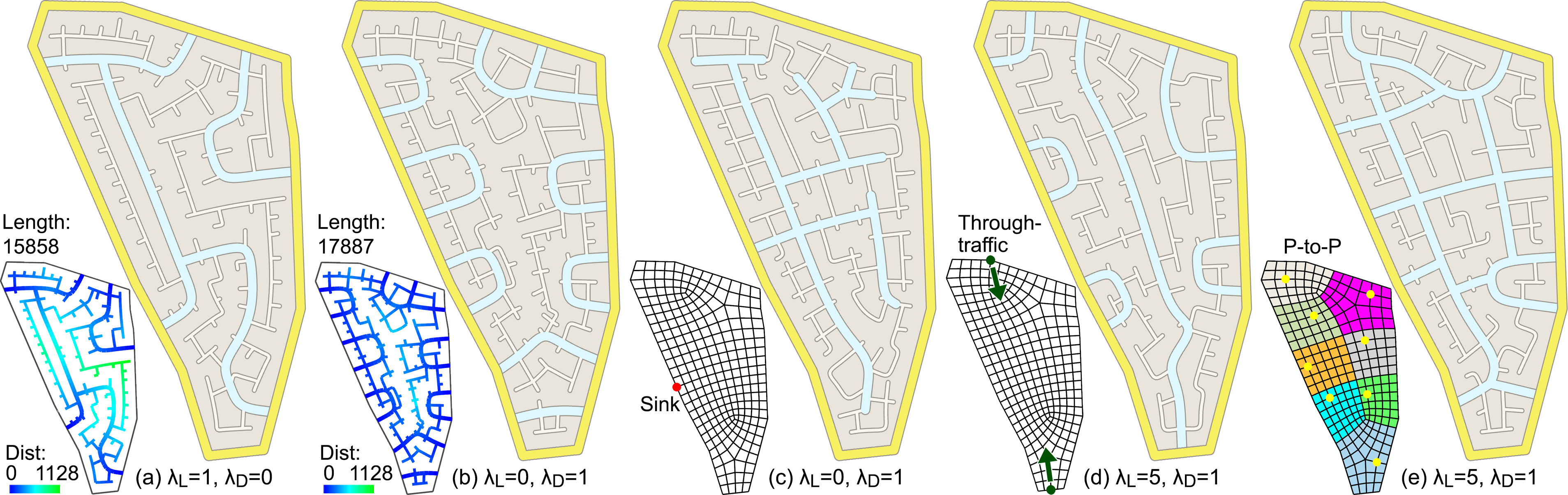}
\caption{Diverse street layouts resulting from different functional specifications. The first two layouts are optimized for (a) minimal network lengths and (b) minimal travel distances to the boundary, using different specifications of the optimization weights. The travel distances are shown in the bottom-left corners. (c) A layout with a single exit on the left. We also prefer a tree-like structure for this case, which is realized by allowing dead-ends on the second (collector roads) level. (d) A layout that encourages through-traffic in the vertical direction. This is realized by enforcing a shortest path connecting the two user-specified vertices (green) without inner branches on the second level. Note that through-traffic in other directions (e.g., horizontal) are implicitly discouraged. (e) A network that better supports interior-to-interior traffic, realized by the point-to-point constraint with a user-specified partition.}
\label{fig:results_plainview}
\end{figure*}

\begin{figure}[b!]
\centering
\includegraphics[width=\linewidth]{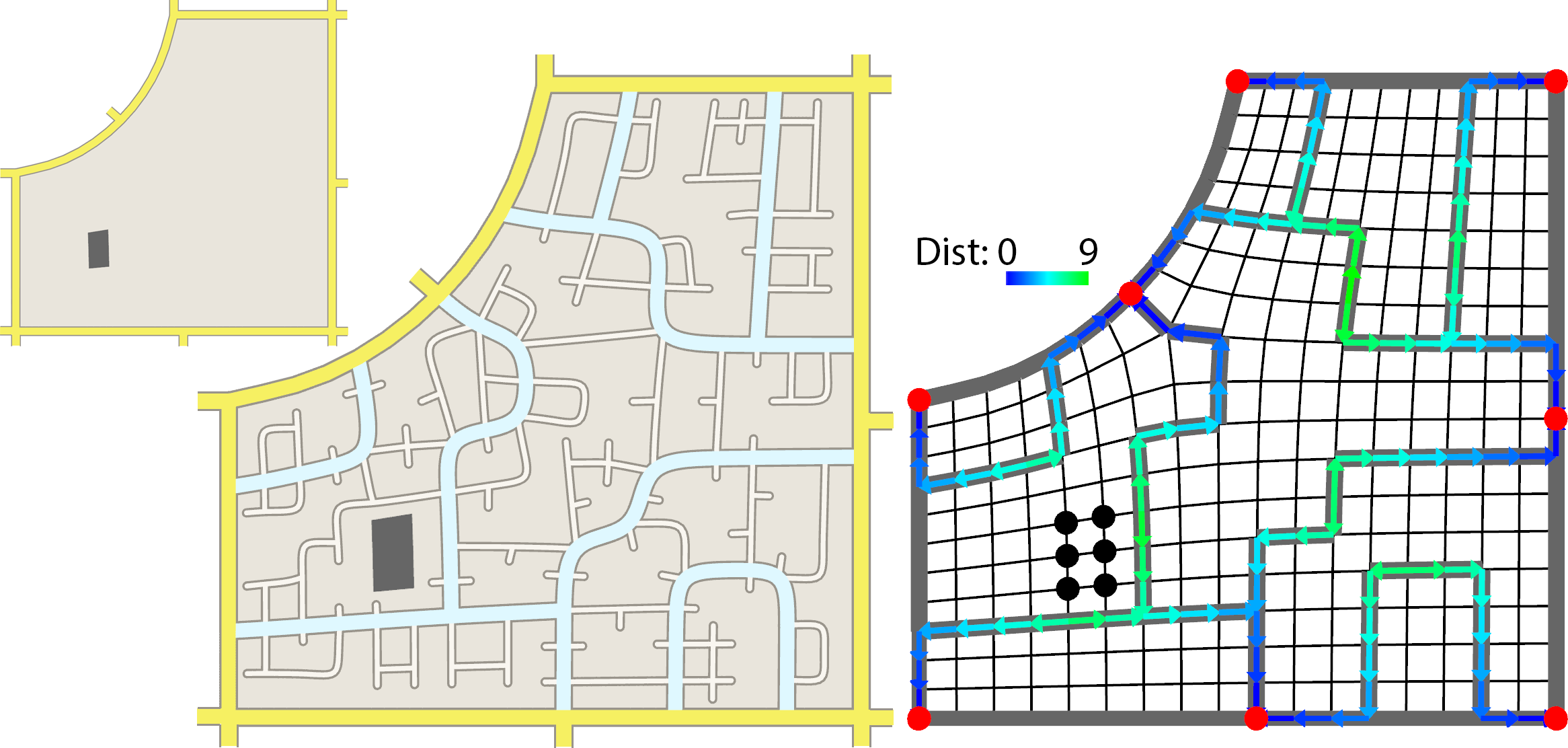}
\caption{Planning a street layout for an empty land surrounded by existing streets and with a historic site that should be preserved. To optimize travel times, instead of designating all boundary vertices as sinks, we place sinks only on the intersections of the existing streets (red vertices). The historic site is preserved by marking the corresponding vertices (black) as obstacles. On the right, we show the IP result of the level-2 roads. The distribution of the active half-edges indicate the shortest paths toward the intersections (sinks) while the distance values encode the shortest distances.}
\label{fig:results_pentagon}
\end{figure}

\begin{enumerate}
\item Density. The density of the street network is directly controlled by the coverage range of the IP.
\item Network lengths versus travel distances. These two competing requirements are controlled by the weights for the total length term ($\lambda_L$) and the total travel distance term ($\lambda_D$) of the IP's objective function.
\item Traffic types (i.e., interior-to-boundary, interior-to-interior, and boundary-to-boundary traffic). By default, we assume that all boundary vertices of the domain mesh are designated as sinks for the IP. This naturally leads to networks that cater to interior-to-boundary traffic while implicitly discouraging traffic of the other two types (Figure~\ref{fig:results_plainview}a and b). To specifically encourage interior-to-interior traffic, the point-to-point constraint of the IP can be used (Figure~\ref{fig:results_plainview}e). To specifically encourage boundary-to-boundary traffic (i.e., through-traffic) in certain directions, simply enforces a shortest path without inner branches in the desired direction (Figure~\ref{fig:results_plainview}d).
\item Sink locations. The overall shape of a street network can be controlled by the distribution of the sinks. Designating all boundary vertices as sinks leads to street networks that are roughly omni-directional toward the boundary. Alternatively, we can designate only a subset of the boundary vertices as sinks, which leads to networks that tilt toward the particular sink vertices (see Figure~\ref{fig:results_plainview}c).
\item Local features. The IP formulation offers direct control over the local features of the generated street networks, including dead-ends, branches, T-junctions, and streets that are too close.
\item Obstacles. The user can easily specify certain locations as obstacles (e.g., water, malls, rail tracks) by enforcing the corresponding vertices or edges to be inactive.
\end{enumerate}

In Figure~\ref{fig:results_plainview} and Figure~\ref{fig:teaser}a-c, we show how distinct street networks for the same sub-region can be created by different functional specifications. In Figure~\ref{fig:results_seaside}, we show a city-level result that consists of multiple sub-region layouts tied together by major roads. In Figure~\ref{fig:results_pentagon}, we consider a practical scenario of designing a street layout for an empty land surrounded by existing streets.

\begin{figure*}[t!]
\centering
\subfloat[]{\label{fig:results_templates}\includegraphics[width=0.125\linewidth]{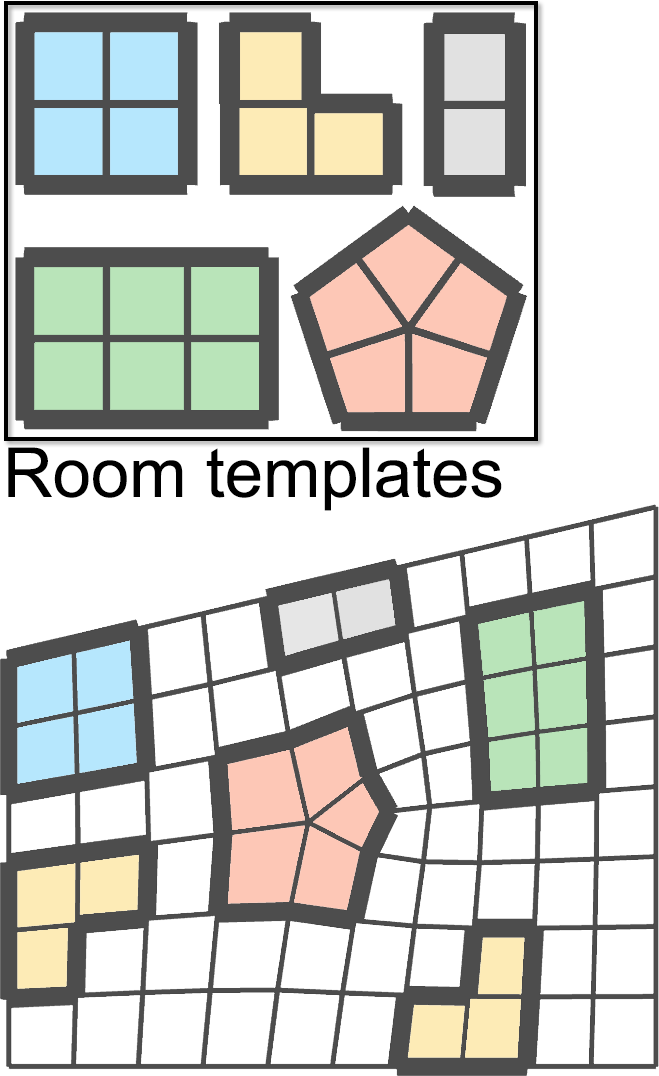}}\quad
\subfloat[]{\label{fig:results_floorplan}\includegraphics[width=0.857\linewidth]{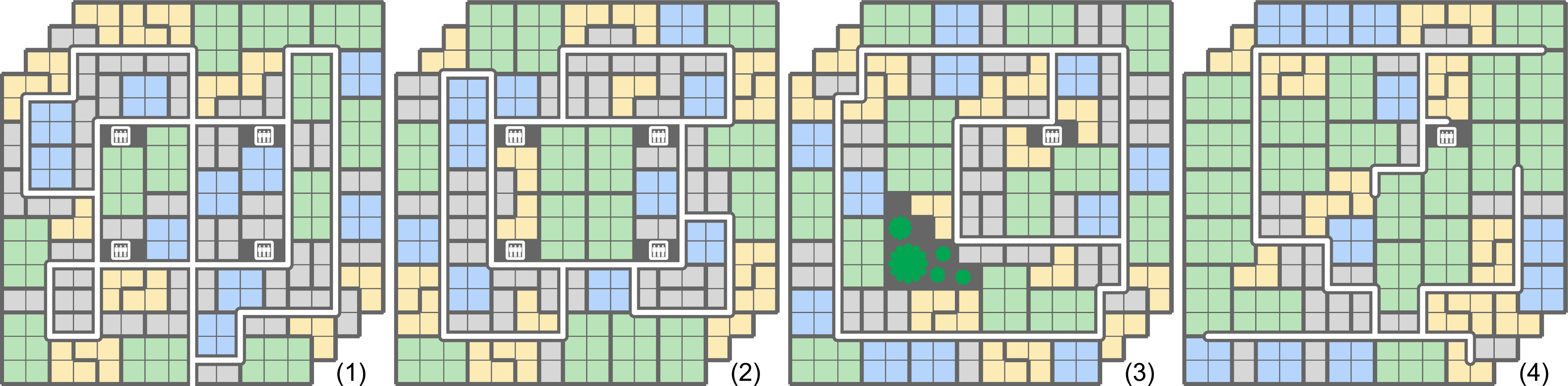}}
\caption{(a) Top: a set of room templates. Bottom: several potential room placements on a mesh. (b) Floor plans for a four-story office building. Note that they consist of both the corridor network and the tiling with room templates. (1) A floor plan with a single sink predefined on the building entrance (bottom middle). The network is constrained to pass through the four elevator locations. Some faces are denoted as obstacles to be occupied by the elevators. (2) A floor plan sharing the same locations of the elevators (as sinks). (3) A floor plan that has a single sink and a large obstacle area for a roof garden. (4) Another floor plan with a single sink. Dead-ends are allowed for the network.}
\end{figure*}

\begin{figure}[t!]
\centering
\includegraphics[width=\linewidth]{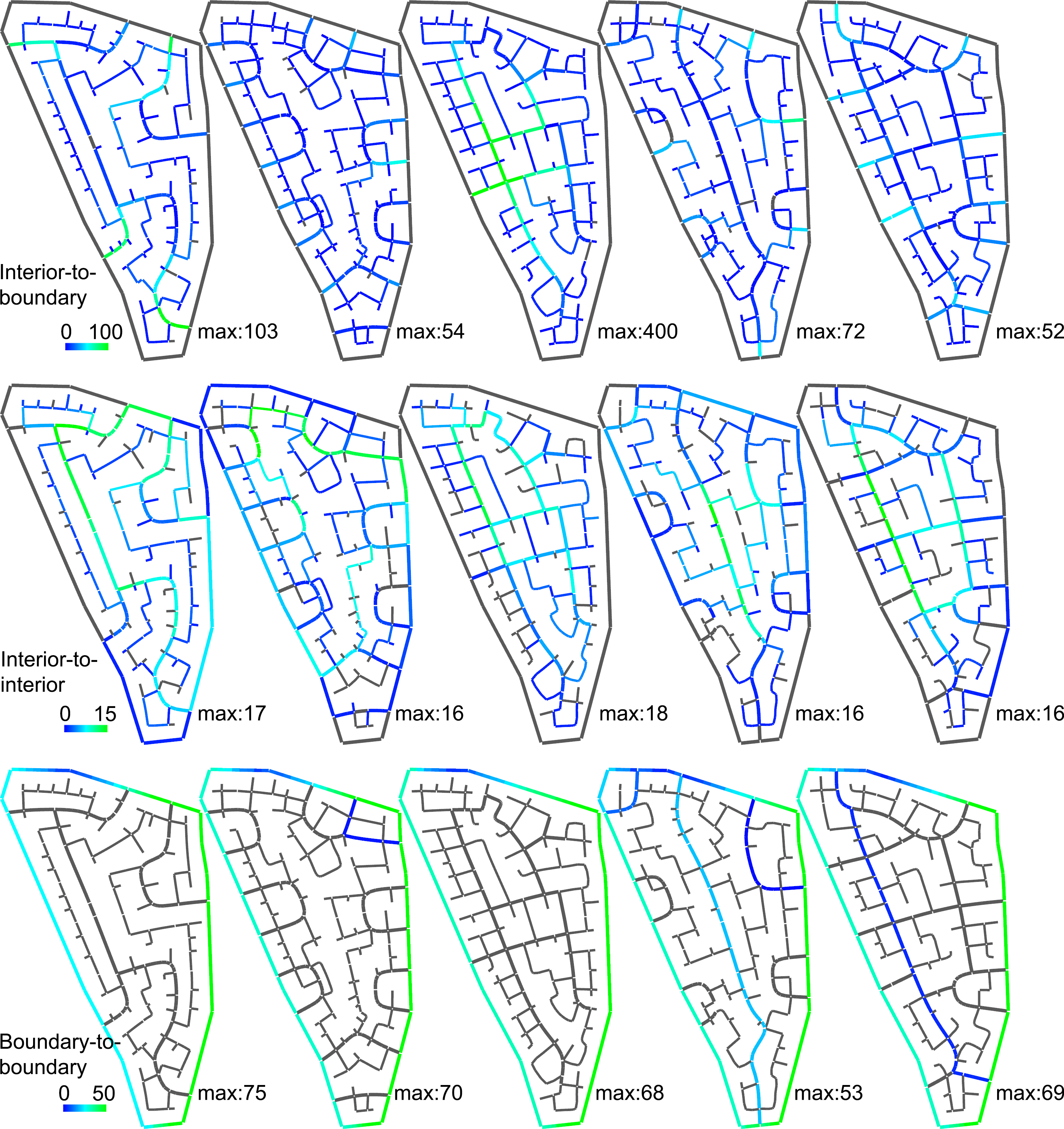}
\caption{SUMO traffic simulations.}
\label{fig:SUMO}
\end{figure}

\mypara{Traffic simulation.} We use SUMO to do traffic simulations to evaluate our functional specifications about the traffic types~\ref{fig:SUMO}.

\subsection{Floorplanning}
\label{sec:floorplan}

Our network generation method is a complement to the tiling-based floorplanning method in Peng et al.~\shortcite{Peng:2014:CLD:2601097.2601164}, of which a major shortcoming is that it is computationally expensive to model the {\em corridors} (i.e., the passage areas connecting the rooms) in a building. In their approach, the corridors are limited to have a tree-like topology, and every level of the tree branches needs a new set of corridor tile templates. Our network IP formulation offers a more general way to model the corridors. We describe our approach next.

We now assume that the given building footprint is discretized into a quad mesh, $M$. A computed network represents the corridors for the building. We replace the coverage constraint by a {\em room tiling} constraint as follows. As detailed in~\cite{Peng:2014:CLD:2601097.2601164}, the user first defines a set of room templates describing the admissible room shapes as combinations of squares, such as a 2x2 square room, a 3x2 long room, and an L-shaped room, with the possibility of having non-valence 4 vertices within (see Figure~\ref{fig:results_templates} top). We then enumerate all possible potential placements of the rooms on $M$. Each potential placement is practically a connected set of faces on $M$ (see Figure~\ref{fig:results_templates} bottom). 

Our goal is to find a subset of all the possible potential room placements such that no two overlap and together they fully cover $M$'s faces. We denote the presences of each potential room placement in the subset as Boolean indicator variables $R_x$, for $x \in [0,N_r)$ with $N_r$ being the number of all potential placements. It follows that:
For every face on $M$, $f$,
\begin{equation}
\sum_i R^{cover}_{i} = 1,
\label{equ:result_tiling0}
\end{equation}
where $R^{cover}_{i}$, $i = 0, 1, ..., X_6 -1$, denotes the set of indicator variables of the potential room placements that cover $f$. For faces denoted as {\em obstacles}, we change the right-hand side of the equation to zero.

In addition, a room has to be connected to the corridors (i.e., active edges of the network) and cannot have corridors in its interior, modeled as follows:
For every potential room placement, $R_x$,
\begin{equation}
X_7 R_x - \sum_{0 \le i < X_7} (1-E_i) \le 0,
\label{equ:result_tiling1}
\end{equation}
where $E_i$, $i = 0, 1, ..., X_7 -1$, denotes the set of indicator variables of the inner edges of $R_x$, and,

\begin{equation}
R_x - \sum_{0 \le j < X_8} E_j \le 0,
\label{equ:result_tiling2}
\end{equation}
where $E_j$, $j = 0, 1, ..., X_8 -1$, denotes the set of indicator variables of the boundary edges of $R_x$.

In summary, the IP formulation for floorplanning comprises of the original network IP formulation (see end of Section~\ref{sec:IP}) but with the coverage constraints (Equation~\ref{equ:IP_coverage0} - \ref{equ:IP_coverage2}) replaced by the room tiling constraints (Equations~\ref{equ:result_tiling0} - \ref{equ:result_tiling2}).

Our floorplanning approach inherits all the functional specifications for modeling networks/corridors (see Section~\ref{sec:urban}) except that the density aspect is now determined by the user-specified admissible room shapes. In addition, as the presences of rooms are explicitly expressed as Boolean variables, users can precisely control the occurrences of each room type using linear constraints. In Figure~\ref{fig:results_floorplan}, we show several floor plans for an office building with distinct functions. In Figure~\ref{fig:teaser}d, we show a floor plan for a large facility.

\subsection{Game Level Design}
\label{sec:game}

\begin{figure}[b!]
\centering
\includegraphics[width=\linewidth]{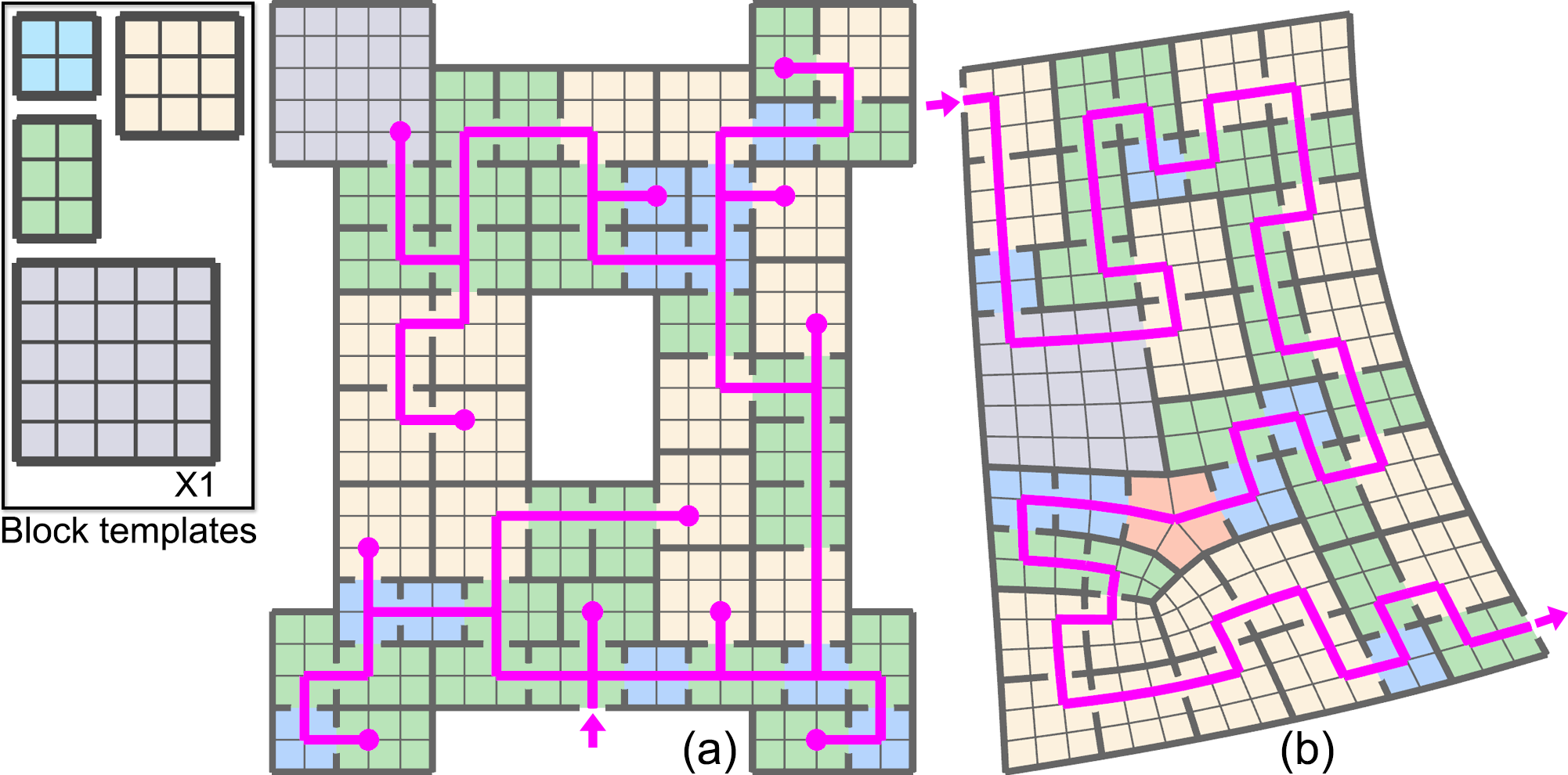}
\caption{Game level designs. The templates for blocks are shown on the left. We constrain the largest room, which encodes a special encounter, to appear exactly once. (a) A design with a tree-like connectivity graph of the blocks. The root is constrained at the castle entrance (lower middle), realized by designating the corresponding vertex to be the sole sink. The player starts at the entrance and explores the whole level in a multiple-choice manner. (b) A design with a linear connectivity graph from the upper-left corner to the bottom-right corner. The player explores every block in the level in a one-by-one manner.}
\label{fig:results_gamelevel}
\end{figure}

We aim to solve one interesting problem in game level design: how to partition a problem domain into {\em blocks} (e.g., rooms or caves) such that the ways the blocks are connected (i.e., their {\em connectivity graphs}) are constrained to ensure various aspects such as difficulty and playability? The problem may come with additional requirements that makes it difficult, such as: (i)~there is a limited number of admissible shapes for blocks; (ii)~ the problem domain has a fixed boundary (e.g., a castle or a dungeon) and the space needs to be fully utilized; and (iii)~the connectivity graph needs to satisfy certain requirements. For example, obviously, all the blocks need to be reachable from certain starting points. Or, whether branches or dead-ends are allowed. Here, we propose an IP-based approach that jointly solves a connectivity graph and a configuration of the building blocks that completely fill the problem domain.

We start from the IP formulation for floorplanning (Section~\ref{sec:floorplan}). The rooms are interpreted as the blocks in a game level in a similar sense. However, we take a different interpretation of the networks: instead of presenting the corridors in a building, we now interpret networks as the the ways the player can traverse the blocks. As seen in Figure~\ref{fig:results_gamelevel}, the network can be understood as a geometric realization of the connectivity graph of the building blocks (assuming that a block is traversed by at most one connected part of the network). This is realized by replacing the floorplanning's constraints about the relationships between rooms and corridors (Equation~\ref{equ:result_tiling1} and~\ref{equ:result_tiling2}) by the following constraints:

For a room (i.e., block) to appear in a game level, at least one of its inner edges need to be active, and all of its boundary edges need to be inactive. That is,

For every potential room placement, $R_x$,
\begin{equation}
R_x - \sum_{0 \le i < X_9} E_i \le 0,
\label{equ:result_game0}
\end{equation}
where $E_i$, $i = 0, 1, ..., X_9-1$, denotes the set of indicator variables of the inner edges of $R_x$, and,

\begin{equation}
X_{10} R_x - \sum_{0 \le j < X_{10}} (1-E_j) \le 0,
\label{equ:result_game1}
\end{equation}
where $E_j$, $j = 0, 1, ..., X_{10} -1$, denotes the set of indicator variables of the boundary edges of $R_x$.

Based on this similar IP formulation, our game level approach has the same functional specifications as our floorplanning approach. In the context of game level design, we can constrain the game level type to be: (i)~{\em branching}, by allowing branches in the connectivity graph (Figure~\ref{fig:results_gamelevel}a), (ii)~{\em circular}, by forbidding branches and designating a single sink (Figure~\ref{fig:teaser}e), or (iii)~{\em linear}, by forbidding branches and designating two sinks on the boundary (Figure~\ref{fig:results_gamelevel}b).

\begin{figure}[h]
\centering
\includegraphics[width=0.75\linewidth]{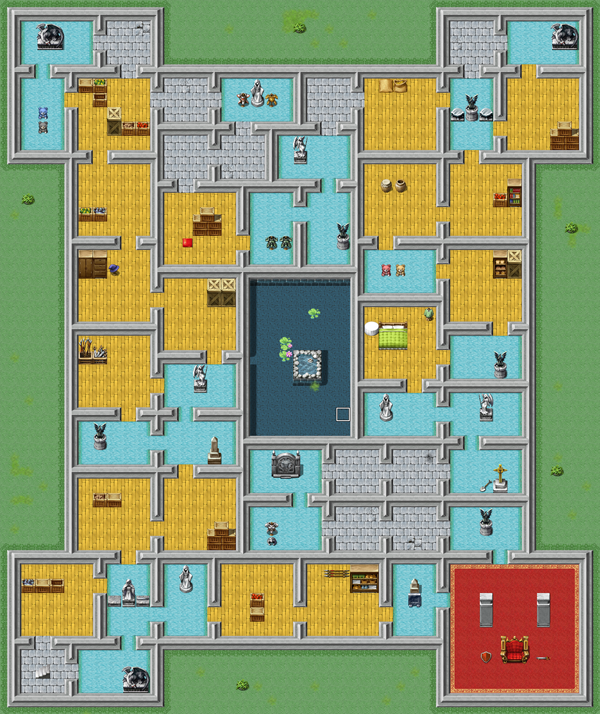}
\caption{We use RPG maker~\protect\cite{RPG} to turn the game level design in Figure~\ref{fig:teaser}e into an actual, playable game level. Each room type is given a different function, e.g., 2x2 blocks are for passages, 3x2 blocks are for enemy encounters, 3x3 blocks are for supplies, and the single 5x5 block is for a boss level. We use additional furniture to separate the blocks that are traversed by more than one connected parts of the connectivity graph.}
\label{fig:results_RPGmaker}
\end{figure}

To demonstrate the usability of our approach in game level design, we use RPG Maker~\cite{RPG} to create actual, playable game levels based on our results. One example is shown in Figure~\ref{fig:results_RPGmaker}.

\subsection{Timing and Analysis}
\label{sec:timing}

We implemented our algorithms using C++ and report timings on a desktop computer with a 2.4 GHz eight-core CPU and 8 GB memory. We use Gurobi~\shortcite{gurobi} to solve the IP problems. The timing statistics (except for the city-level result) are shown in Table~\ref{table:timing}. In practice, as it is difficult to find a global optimum, we also accept sub-optimal solutions (fulfilling all hard constraints) computed within reasonable time limits.

The IP can become infeasible due to hard constraints - for example, it would apparently become infeasible if the coverage constraint requires that all vertices need to be covered and edges being too close are forbidden. However, the Gurobi solver promptly identifies a problem as infeasible.

\begin{table}[!h]
\caption{For every example shown in the paper, we show the number of edges in the mesh, the parameters for the IP, and the times to obtain the shown solutions. inf (i.e., infinite) means the corresponding features is forbidden. Y means the feature is allowed and N means forbidden. For urban street layouts, the times to calculate the results of level-2, level-3, and level-4 are shown.}
\small
\begin{tabular}{|p{0.06\linewidth}p{0.04\linewidth}p{0.06\linewidth}|p{0.22	\linewidth}p{0.03\linewidth}p{0.04\linewidth}p{0.03\linewidth}|p{0.12\linewidth}|}\hline

& Mesh & & $\lambda_L$, $\lambda_D$, & Dead & & & Time \\
& edge & vars&  $\lambda^0_Z$, $\lambda^1_Z$, $\lambda_T$ & end & Branch & PtP & (seconds) \\ \hline

Fig~\ref{fig:teaser}a & 657 & 6092$\dagger$ & 5 , 1 , 5 , inf , 0 & N & Y & N & 379/366/49\\
Fig~\ref{fig:teaser}b & 657 & 7470$\dagger$ & 0 , 1 , 5 , inf , 10$\ddagger$ & N & Y & Y & 168/274/34\\
Fig~\ref{fig:teaser}c & 657 & 6231$\dagger$ & 1 , 0 , 5 , inf , 0 & N & Y & N & 456/467/-\\
Fig~\ref{fig:teaser}d & 1012 & 11338 & 1 , 0 , 5 , inf , 0 & N & Y & N & 920\\
Fig~\ref{fig:teaser}e & 782 & 12056 & 1 , 0 , 5 , inf , 0 & N & N & N & 4927\\
Fig~\ref{fig:IP_results}a & 312 & 2241 & 1 , 0* , 5 , inf , 0 & N & Y & N & 132\\
Fig~\ref{fig:IP_results}b & 312 & 2241 & 0 , 1 , 5 , inf , 0 & N & Y & N & 35\\
Fig~\ref{fig:IP_results}c & 312 & 2281 & 0 , 1 , 5 , inf , 0 & N & Y & Y & 11\\
Fig~\ref{fig:IP_results}d & 312 & 2766 & 0 , 1 , 5 , inf , 0 & N & Y & N & 24\\
Fig~\ref{fig:IP_results}e & 312 & 3865 & 1 , 0* , 5 , inf , 0 & N & Y & N & 74\\
Fig~\ref{fig:IP_results}f & 312 & 3865 & 0 , 1 , 5 , inf , 0 & N & Y & N & 56\\
Fig~\ref{fig:IP_results}g & 312 & 4751 & 0 , 1 , 5 , inf , 0 & N & Y & Y & 26\\
Fig~\ref{fig:results_pipeline}L3 & 1200 & 6960 & 1 , 0 , 5 , inf , 0 & N & Y & N & 86\\
Fig~\ref{fig:results_pipeline}L4 & 1200 & 1930 & 1 , 0 , 5 , inf , 0 & Y & Y & N & 2\\
Fig~\ref{fig:results_plainview}a & 535 & 4432 & 1 , 0* , inf , inf , 0 & N & Y & N & 121/218/23\\
Fig~\ref{fig:results_plainview}b & 535 & 4432 & 0 , 1 , 5 , inf , 0 & N & Y & N & 174/318/76\\
Fig~\ref{fig:results_plainview}c & 535 & 5304 & 0 , 1 , 5 , inf , 0 & Y & Y & N & 70/294/46\\
Fig~\ref{fig:results_plainview}d & 535 & 5036 & 5 , 1 , 5 , inf , 0 & N & Y & N & 12/223/43\\
Fig~\ref{fig:results_plainview}e & 535 & 5651 & 5 , 1 , 5 , inf , 0 & N & Y & Y & 73/195/114\\
Fig~\ref{fig:results_pentagon} & 543 & 4415 & 0 , 1 , 5 , inf , 0 & N & Y & N & 229/500/1\\
Fig~\ref{fig:results_floorplan}1 & 520 & 7986 & 1 , 0 , 5 , inf , 0 & N & Y & N & 1018\\
Fig~\ref{fig:results_floorplan}2 & 520 & 7876 & 1 , 0 , 5 , inf , 0 & N & Y & Y & 188\\
Fig~\ref{fig:results_floorplan}3 & 520 & 7944 & 1 , 0 , 5 , inf , 0 & N & Y & N & 201\\
Fig~\ref{fig:results_floorplan}4 & 520 & 7944 & 1 , 0 , 5 , inf , 0 & Y & Y & N & 1507\\
Fig~\ref{fig:results_gamelevel}a & 782 & 11411 & 1 , 0 , 5 , inf , 0 & Y & Y & N & 3359\\
Fig~\ref{fig:results_gamelevel}b & 657 & 9643 & 1 , 0 , 5 , inf , 0 & N & N & N & 2958\\
\hline		
\end{tabular}
*: 1e-4. $\dagger$:L1. $\ddagger$:L4. \\
\label{table:timing}
\end{table}

For the city-level result (Figure~\ref{fig:results_seaside}), the statistics for the sub-regions are shown in the additional material. The sum of computation times is 7844 seconds. However, as the computations for each sub-region are independent, they can be done in parallel. When done in parallel, it takes about 2300 seconds (using a time limit of 1000 seconds for level 2, 1200 seconds for level 3, and 100 seconds for level 4) to compute a comparable result.

The main limitation is performance. As it usually takes a few minutes to solve a medium-sized urban layout problem, interactive speed is not yet achieved. A restriction of the IP formulation is that new additions/modifications should be linear, otherwise, the problem becomes too expensive to solve.

\subsection{Comparisons with Other Approaches}
\label{sec:comparison}

In this section, we show that it is advantageous in terms of performance to formulate network problems into IP form and solve with a specialized IP solver (e.g., Gurobi).

\begin{figure}[h!]
\centering
\includegraphics[width=\linewidth]{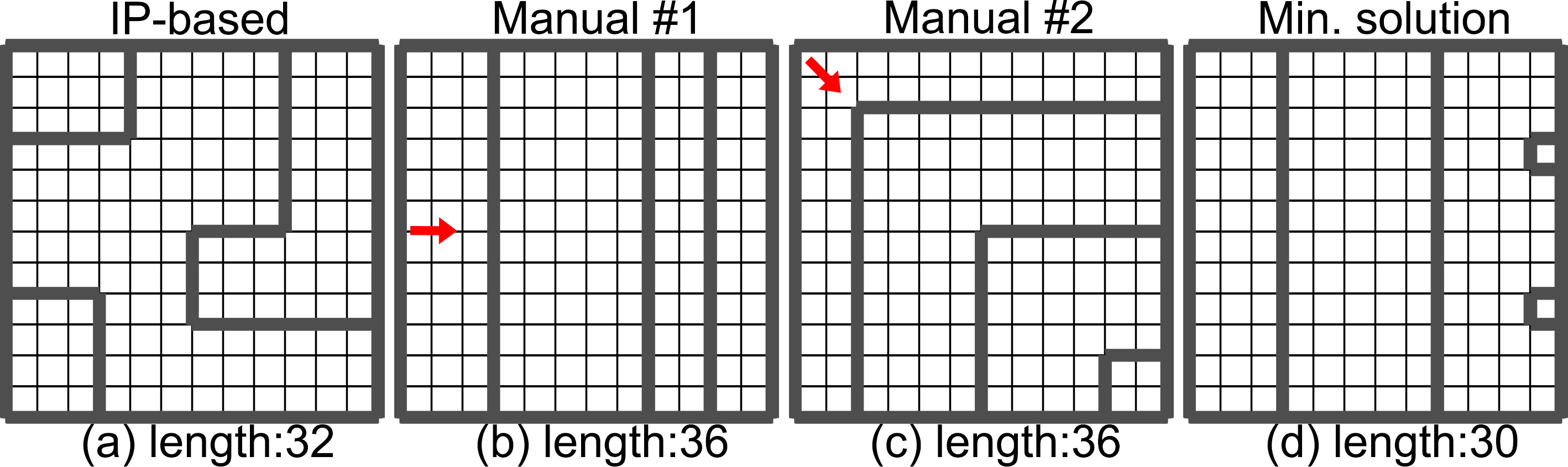}
\caption{(a) Our IP-based solution to find a network that cover the mesh (coverage range is two edges wide) with the fewest possible number of edges while avoiding zig-zags and edges that are too close. (b) and (c) Two manual results for comparisons. Our strategy is to start at one side of the boundary or a corner and grow edges as far as possible. Zig-zags and edges that are too close are avoided. (d) An optimal solution that allows zig-zags and edges that are too close, found by our IP approach.}
\label{fig:results_trivial}
\end{figure}

\mypara{Manual solution} We first compare our solutions to some trivial solutions created manually. In Figure~\ref{fig:results_trivial}, We manually create solutions to achieve the same optimization goals as in Figure~\ref{fig:IP_results}a. Such solutions use more edges than our IP-based solution. We also attempted to create floorplanning results by hand. We find that it is very challenging to create full room tilings manually, let alone jointly finds a valid network that satisfies the given constraints.

\begin{figure}[h!]
\centering
\includegraphics[width=\linewidth]{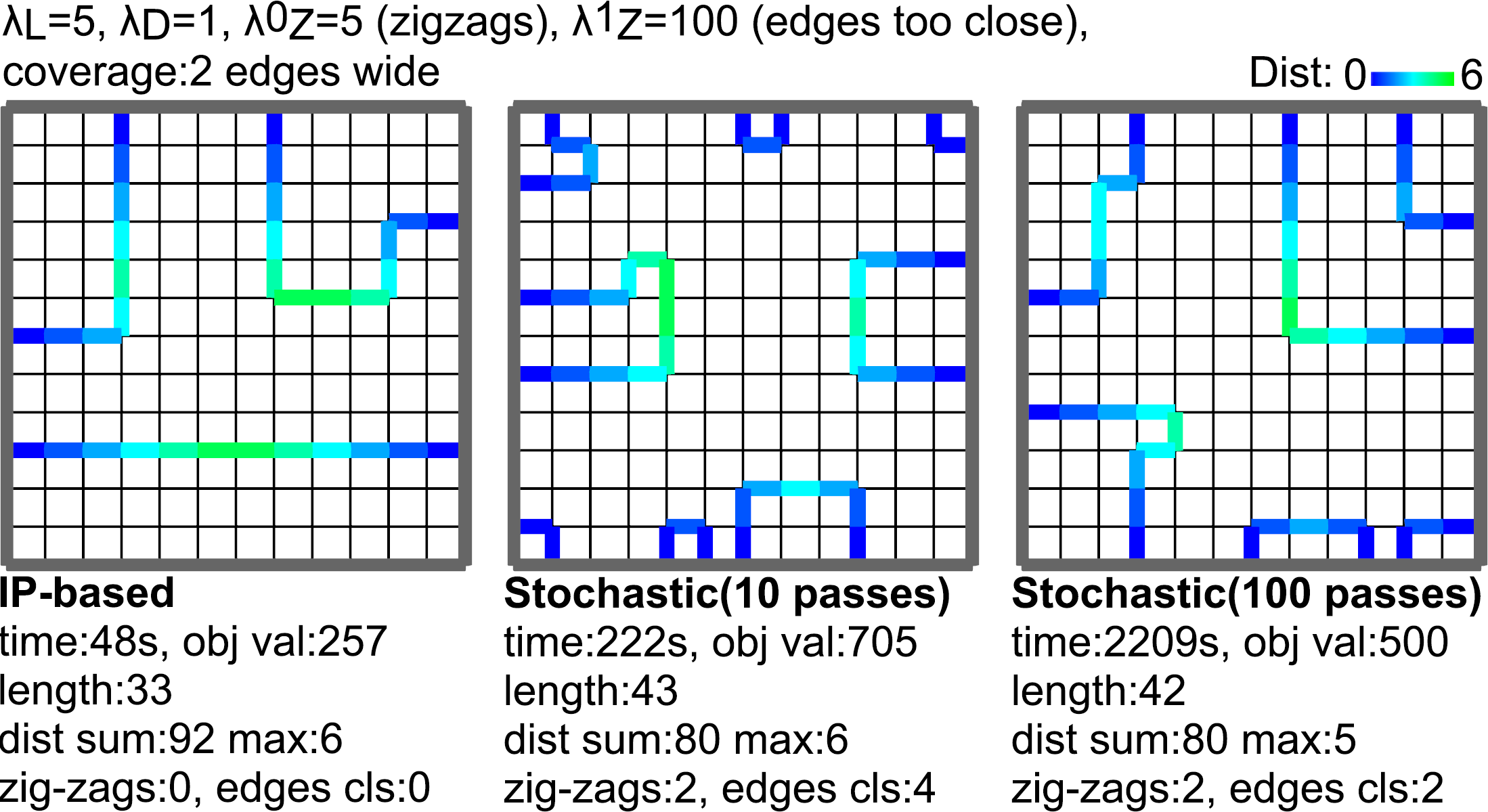}
\caption{Comparing to a stochastic search-based approach. We run the approach multiple passes and pick the best solution. Even with a much longer time, the stochastic approach cannot find solutions of comparable qualities.}
\label{fig:results_stochastic}
\end{figure}

\mypara{Stochastic search} For comparison, we implemented a stochastic search-based approach to solve the network problems. Beginning at a trivial solution (e.g., every edge is active), the approach iteratively performs the following types of operations to alter the current solution: (i)~deleting a single edge, (ii)~deleting a pair of adjacent edges, (iii) deleting a triple of consecutive edges, and (iv)~adding a single edge. At each iteration, we enumerate all possible feasible operations, rank them according to the new objective values (the lower the better), and pick one to perform. We pick operations in a simulated annealing sense (i.e., the higher ranked the larger chance to be picked, and such tendency becomes more absolute at each iteration). The approach stops when there are no feasible solutions or a time limit is reached. As shown in Figure~\ref{fig:results_stochastic}, we find that such a stochastic approach cannot compete with the IP-based approach in terms of speed and result qualities.

\section{Conclusions and Future Work}

We proposed an algorithm for the computational design of networks for layout computation, such as street networks, building floor plans, and game levels. The user provides high-level functional specifications for the target problem domain, while our algorithm jointly realizes the connectivity and the detailed geometry of the network. While there is a considerable amount of work on using functional specifications for evaluating networks, to the best of our knowledge, this is the first attempt to synthesize these layouts {\em purely} based on functional specifications.

In future work, it is interesting to consider multi-modal transportation networks (e.g., public transportations) for a richer variety of urban street layouts. We would also like to tackle other network design problems by our IP-based approach, such as the layouts of residential houses, automated warehouses, and electrical layouts. In addition, while the meshes used in this paper are all quadrilateral because of our target applications, new design problems may necessitate the need for more kinds of mesh tessellations, such as a hybrid of quad and triangle meshes.



\bibliographystyle{acmsiggraph}
\bibliography{functional_layout}

\section*{Appendix}
\label{sec:smoothing}

\mypara{Snake-based smoothing} We use the active contour model (snakes)~\cite{Kass:1988:SACM} to smooth the coarse street networks generated by the IP approach, which tend to contain many sharp turns (e.g., stair-shaped) due to the nature of quad meshes. We give a summary of the algorithm as follows.

A snake is a distinct non-empty sequence of active edges that connects a distinct sequence of vertices (i.e., no branches nor loops). Moreover, the valence of the first and last vertices of a snake cannot be $2$; that is, a snake must end at non valence-$2$ vertices. We first decompose a street network into snakes. It is straightforward to see that a network can be decomposed into non-overlapping snakes that together fully cover the network. Note that snakes can include intersection vertices of the street network, i.e., an active vertex that connects to more than two active edges, from its interior. After the snakes are extracted, we subdivide each snake so that the smoothing algorithm, described next, may have a higher degrees of freedom.

The snake-based smoothing algorithm minimizes the energy associated with each snake, which can be understood as deformable splines. We consider three types of energies: tension, stiffness, and inertia. Assuming that a snake is represented by $\mathbf{u}(s, t)$, where $s \in (0, 1)$ is the parametric domain and $t$ is the time (i.e., iteration), for each vertex, the energies are defined as follows.

\begin{itemize}
  \item Tension energy, $E_\text{tension}(\mathbf{u})=\left| \frac{\partial \mathbf{u}(s)}{\partial s} \right|^2$. Minimizing the tension energy makes the snake act like a membrane. A higher weight for this energy leads to shorter lengths.
  \item Stiffness energy, $E_\text{stiffness}(\mathbf{u})=\left| \frac{\partial^2 \mathbf{u}(s)}{\partial^2 s} \right|^2$. Minimizing the stiffness energy makes the snake act like a thin plate. A higher weight for this energy leads to smoother turns.
  \item Inertia energy, $E_\text{inertia}(\mathbf{u})=\left| \mathbf{u}(s, t) - \mathbf{u}(s, 0) \right|^2$. This energy is used to prevent the snake from moving too far away from its original position.
\end{itemize}

The overall energy is defined as a linear combination of these three energies for all vertices:
\begin{equation}
E= \sum_\text{snake} \int_0^1 \left(\alpha E_\text{tension} + \beta E_\text{stiffness} + \gamma E_\text{inertia}\right) \mathrm{d}s.
\end{equation}
Here, $\alpha$, $\beta$, and $\gamma$ are the weights of the three energies.

We use different settings of weights at different levels of the street layout hierarchy. Streets at higher levels use higher stiffness weights (smoother corners) while streets at lower levels use higher inertia weights (fewer deformations). Additional heuristics are used to make sure the street intersections are close to 90$^\circ$ angles.

We use gradient descent to optimize the snake energy. The gradient of the snake energy is:
\begin{equation}
\begin{split}
\nabla E =\sum_\text{snake} \int_0^1 \bigg(-2\alpha \frac{\partial^2 \mathbf{u}(s)}{\partial s^2} + 2\beta \frac{\partial^4 \mathbf{u}(s)}{\partial s^4}+ \\
 2\gamma \left(\mathbf{u}(s, t) - \mathbf{u}(s, 0) \right) \bigg) \mathrm{d}s.
\end{split}
\end{equation}
We take adaptive steps in the direction of $-\nabla E$ until the changes stabilize.

\end{document}